\documentstyle[epsf,12pt,amstex]{article}
\def\baselinestretch{1.1}
\newtheorem{theorem}{Theorem}

\numberwithin{equation}{section}

\begin{document}
\title{\bf \huge Multifractal characterisation of complete genomes}
\author{Vo Anh$^1$, Ka-Sing Lau$^2$ and Zu-Guo Yu$^{1,3}$\\
{\small $^1$Centre in Statistical Science and Industrial Mathematics, Queensland
University}\\
{\small  of Technology, GPO Box 2434, Brisbane, Q4001, Australia}\\
{\small $^2$Department of Mathematics, Chinese University of Hong Kong, 
Shatin, Hong Kong}\\
{\small $^3$Department of Mathematics, Xiangtan University, Hunan 411105, P.R. China}
}
\date{}
\maketitle

\begin{abstract}
This paper develops a theory for characterisation of DNA sequences based on
their measure representation. The measures are shown to be random cascades
generated by an infinitely divisible distribution. This probability
distribution is uniquely determined by the exponent function in the
multifractal theory of random cascades. Curve fitting to a large number of
complete genomes of bacteria indicates that the Gamma density function provides an
excellent fit to the exponent function, and hence to the probability
distribution of the complete genomes.
\end{abstract}

\footnotetext[1]{ E-mail address: v.anh@@qut.edu.au, kslau@@math.cuhk.edu.hk, 
yuzg@@hotmail.com or z.yu@@qut.edu.au}
\footnotetext[2]{ The URL of Vo Anh: http://www.maths.qut.edu.au/cissaim/anh.html}

\section{Introduction}

\ \ \ \ DNA sequences are of fundamental importance in understanding living
organisms, since all information on their hereditary evolution is contained
in these macromolecules. One of the challenges of DNA sequence analysis is
to determine the patterns of these sequences. It is useful to distinguish
coding from noncoding sequences. Problems related to the classification and
evolution of organisms using DNA sequences are also important.

Fractal analysis has proved useful in revealing complex patterns in natural
objects. Berthelsen \textit{et al.} \cite{BGR94} considered the global
fractal dimension of human DNA sequences treated as pseudorandom walks.
Vieira \cite{Vie99} carried out a low-frequency analysis of the complete DNA
of 13 microbial genomes and showed that their fractal behaviour does not
always prevail through the entire chain and the autocorrelation functions
have a rich variety of behaviours including the presence of
anti-persistence. Provata and Almirantis \cite{PA00} proposed a fractal
Cantor pattern of DNA. They mapped coding segments to filled regions and
noncoding segments to empty regions of a random Cantor set and then
calculated the fractal dimension of this set. They found that the
coding/noncoding partition in DNA sequences of lower organisms is
homogeneous-like, while in the higher eucariotes the partition is fractal.
Yu and Anh \cite{YA00} proposed a time series model based on the global
structure of the complete genome and found that one can get more information
from this model than that of fractal Cantor pattern. Some results on the
classification and evolution relationship of bacteria were found in \cite
{YA00}. The correlation property of length sequences was discussed in \cite
{YAW00}.

Although statistical analysis performed directly on DNA sequences has
yielded some success, there has been some indication that this method is not
powerful enough to amplify the difference between a DNA sequence and a
random sequence as well as to distinguish DNA sequences themselves in more
details. One needs more powerful global and visual methods. For this
purpose, Hao \textit{et al. }\cite{HLZ00} proposed a visualisation method
based on coarse-graining and counting of the frequency of appearance and
strings of a given length. They called it the \textit{portrait} of an
organism. They found that there exist some fractal patterns in the portraits
which are induced by avoiding and under-represented strings. The fractal
dimension of the limit set of portraits was discussed in \cite{YHXC00,HXYC01}%
. There are other graphical methods of sequence patterns, such as chaos game
representation (see \cite{Jef90, Gol93}).

In the portrait representation, Hao \textit{et al. }\cite{HLZ00} used
squares to represent substrings and discrete colour grades to represent the
frequencies of the substrings in the complete genome. It is difficult to
know the accurate value of the frequencies of the substrings from the
portrait representation. And they did not discuss the classification and
evolution problem. In order to improve it, Yu {\it et al} \cite{YA00b}
used subintervals in one-dimensional space to represent substrings to obtain
an accurate histogram of the substrings in the complete genome. The
histogram, viewed as a probability measure and was called the \textit{%
measure representation }of the complete genome, gives a precise compression
of the genome. Multifractal analysis was then proposed in Yu {\it et al} \cite
{YA00b} to treat the classification and evolution problem based on the
measure representation of different organisms.

In this paper, we go one step further and provide a complete
characterisation of the DNA sequences based on their measure representation.
This is given in the form of the probability density function of the
measure. We first show that the given measure is in fact a multiplicative
cascade generated by an infinitely divisible distribution. This probability
distribution is uniquely determined by the exponent $K\left( q\right) ,$ $%
q\geq 0,$ in the multifractal analysis of the cascade. This theory will be
detailed in the next section. We then apply the theory on a large number of
typical genomes. It will be seen that the Gamma density function provides an
excellent fit to the $K\left( q\right) $ curve of each genome. This
characterisation therefore provides a needed tool to study the evolution of
organisms.

\section{Measure representation of complete genome}

\ \ \ \ We first outline the method of Yu {\it et al} \cite{YA00b} in deriving the
measure representation of a DNA sequence. Such a sequence is formed by four
different nucleotides, namely adenine ($a$), cytosine (\emph{c)}, guanine (%
\emph{g)} and thymine (\emph{t). }We call any string made of $K$ letters
from the set $\{g,c,a,t\}$ a $K$-string. For a given $K$ there are in total $%
4^{K}$ different $K$-strings. In order to count the number of each kind of $%
K $-strings in a given DNA sequence, $4^{K}$ counters are needed. We divide
the interval $[0,1[$ into $4^{K}$ disjoint subintervals, and use each
subinterval to represent a counter. Letting $s=s_{1}\cdots s_{K},s_{i}\in
\{a,c,g,t\},i=1,\cdots ,K,$ be a substring with length $K$, we define 
\begin{equation}
x_{l}(s)=\sum_{i=1}^{K}\frac{x_{i}}{4^{i}},
\end{equation}
where 
\begin{equation}
x_{i}=\left\{ 
\begin{array}{l}
0,\ \ \ \mbox{if}\ s_{i}=a, \\ 
1,\ \ \ \mbox{if}\ s_{i}=c, \\ 
2,\ \ \ \mbox{if}\ s_{i}=g, \\ 
3,\ \ \ \mbox{if}\ s_{i}=t,
\end{array}
\right.
\end{equation}
and 
\begin{equation}
x_{r}(s)=x_{l}(s)+\frac{1}{4^{K}}.
\end{equation}
We then use the subinterval $[x_{l}(s),x_{r}(s))$ to represent substring $s$%
. Let $N(s)$ be the times of substring $s$ appearing in the complete genome.
If the number of bases in the complete genome is $L$, we define 
\begin{equation}
F(s)=N(s)/(L-K+1)
\end{equation}
to be the frequency of substring $s$. It follows that $\sum_{\{s\}}F(s)=1$.
Now we can  define a measure $\mu _{K}$
on $[0,1)$ by 
\begin{equation*}
\mu _{K}\left( dx\right) =Y_{K}\left( x\right) dx,
\end{equation*}
where 
\begin{equation}
Y_{K}(x)=4^{K}F_{K}(s),\ \ \ \ x\in \lbrack x_{l}(s),x_{r}(s)).
\end{equation}
We then have $\mu _{K}\left( [0,1)\right) =1$ and $\mu _{K}\left(
[x_{l}(s),x_{r}(s))\right) =F_{K}(s).$ We call $\mu _{K}\left( x\right) $
the \textit{\ measure representation} of an organism. As an example, the
measure representation of \textit{M. genitalium} for $K=3,...,8$ is given in
Figure 1. Self-similarity is apparent in the measures.

More than 33 bacterial complete genomes are now available in public
databases. There are six Archaebacteria (\textit{Archaeoglobus fulgidus}, 
\textit{Pyrococcus abyssi}, \textit{Methanococcus jannaschii}, \textit{%
Pyrococcus horikoshii}, \textit{Aeropyrum pernix} and \textit{%
Methanobacterium thermoautotrophicum)}; five Gram-positive Eubacteria (%
\textit{Mycobacterium tuberculosis}, \textit{Mycoplasma pneumoniae}, \textit{%
Mycoplasma genitalium}, \textit{Ureaplasma urealyticum}, and \textit{%
Bacillus subtilis)}. The others are Gram-negative Eubacteria, which consist
of two Hyperthermophilic bacteria (\textit{Aquifex aeolicus} and \textit{%
Thermotoga maritima)}; five Chlamydia (\textit{Chlamydia trachomatisserovar}%
, \textit{Chlamydia muridarum}, \textit{Chlamydia pneumoniae} and \textit{%
Chlamydia pneumoniae AR39)}; two Spirochaete (\textit{Borrelia burgdorferi}
and \textit{Treponema pallidum)}; one Cyanobacterium (\textit{Synechocystis
sp. PCC6803)}; and thirteen Proteobacteria. The thirteen Proteobacteria are
divided into four subdivisions, which are alpha subdivision (\textit{%
Rhizobium sp. NGR234} and \textit{Rickettsia prowazekii)}; gamma subdivision
(\textit{Escherichia coli}, \textit{Haemophilus influenzae}, \textit{Xylella
fastidiosa}, \textit{Vibrio cholerae}, \textit{Pseudomonas aeruginosa} and 
\textit{Buchnera sp. APS)}; beta subdivision (\textit{Neisseria meningitidis
MC58} and \textit{Neisseria meningitidis Z2491)}; epsilon subdivision (%
\textit{Helicobacter pylori J99}, \textit{Helicobacter pylori 26695} and 
\textit{Campylobacter jejuni)}. The complete sequences of some chromosomes
of higher organisms are also currently available. We selected the sequences
of Chromosome 15 of \textit{Saccharomyces cerevisiae}, Chromosome 3 of 
\textit{Plasmodium falciparum}, Chromosome 1 of \textit{Caenorhabditis
elegans}, Chromosome 2 of \textit{Arabidopsis thaliana} and Chromosome 22 of 
\textit{Homo sapiens}.

In our previous work (Yu {\it et al} \cite{YA00b}), we calculated the numerical
dimension spectra $D_q$ (defined in next section)
for all above organisms and for different $K$. For small $K$, 
there are only
a few different $K$-strings, so there is not enough information 
for any clear-cut result. 
We find that the $D_q$ curves are very close to one another for $K=6,7,8$ for each
organism. Hence
it would be appropriate to take $K=8$ if we want to use the $D_q$ curves to discuss
the classification and evolution problem.  It is still needed
to know what is the analytical expression of the dimension spectra.
The main aim of this paper is to establish a theoretical model to give
such an analytical expression.

\section{Multifractal models}

\ \ \ \ Let $\varepsilon \left( t\right) $ be a positive stationary stochastic
process on a bounded interval of $\mathbb{R},$ assumed to be the unit
interval $\left[ 0,1\right] $ for convenience, with $E\varepsilon \left(
t\right) =1.$ The smoothing of $\varepsilon \left( t\right) $ at scale $r>0$
is defined as 
\begin{equation}
\varepsilon _{r}\left( t\right) =\frac{1}{r}\int_{t-r/2}^{t+r/2}\varepsilon
\left( s\right) ds.  \label{2.1}
\end{equation}
For $0<r<u<v,$ we consider the processes 
\begin{equation*}
X_{r,v}\left( t\right) =\frac{\varepsilon _{r}\left( t\right) }{\varepsilon
_{v}\left( t\right) },\;t\in \left[ 0,1\right] .
\end{equation*}
Following Novikov \cite{Nov94}, we assume the following scale invariance
conditions:

\begin{enumerate}
\item[(i)]  The random variables $X_{r,u}$ and $X_{u,v}$ are independent;

\item[(ii)]  The probability distribution of each random variable $X_{u,v}$
depends only on the ratio $u/v$ of the corresponding scales.
\end{enumerate}

These conditions imply the power-law form for the moments of the processes $%
X_{u,v}$ if they exist. In fact, we may write 
\begin{equation}
E\left( X_{u,v}\left( t\right) \right) ^{q}=g_{q}\left( \frac{u}{v}\right)
,\;q\geq 0  \label{2.2}
\end{equation}
from condition (ii) for some function $g$ which also depends on $q$. From
the identity 
\begin{equation*}
X_{r,v}\left( t\right) =X_{r,u}\left( t\right) X_{u,v}\left( t\right)
\end{equation*}
and condition (i) we get 
\begin{equation}
g_{q}\left( \frac{r}{v}\right) =g_{q}\left( \frac{r}{u}\right) g_{q}\left( 
\frac{u}{v}\right) .  \label{2.3}
\end{equation}
Since $u$ is arbitrary, we then have 
\begin{equation}
g_{q}\left( \frac{r}{v}\right) =\left( \frac{r}{v}\right) ^{-K\left(
q\right) }  \label{2.4}
\end{equation}
for some function $K\left( q\right) $ with $K\left( 0\right) =0.$ It follows
that 
\begin{equation*}
K\left( q\right) =\frac{\ln E\left( X_{r,v}\left( t\right) \right) ^{q}}{\ln
\left( v/r\right) }.
\end{equation*}
Writing $Y$ for $X_{r,v}$ we obtain 
\begin{equation*}
K^{\prime }\left( q\right) =\frac{1}{\ln \left( v/r\right) }\frac{E\left(
Y^{q}\ln Y\right) }{E\left( Y^{q}\right) },
\end{equation*}
\begin{equation*}
K^{\prime \prime }\left( q\right) =\frac{1}{\ln \left( v/r\right) }\frac{%
\left( EY^{q}\right) E\left( Y^{q}\left( \ln Y\right) ^{2}\right) -\left(
E\left( Y^{q}\ln Y\right) \right) ^{2}}{\left( EY^{q}\right) ^{2}}.
\end{equation*}
Since 
\begin{eqnarray}
\left( E\left( Y^{q}\ln Y\right) \right) ^{2} &=&\left( E\left(
Y^{q/2}Y^{q/2}\ln Y\right) \right) ^{2}  \notag \\
&\leq &\left( EY^{q}\right) E\left( Y^{q}\left( \ln Y\right) ^{2}\right)
\label{2.5}
\end{eqnarray}
by Schwarz's inequality and $v/r>1,$ we get $K^{\prime \prime }\left(
q\right) \geq 0,$ that is, $K\left( q\right) $ is a convex function. It is
noted that equality holds in (\ref{2.5}) only if $K\left( q\right) $ is a
linear function of $q;$ other than this, $K\left( q\right) $ is a strictly
convex function.

For $0<q<1,$ we assume that $K\left( q\right) <0,$ which reflects the fact
that, in this range, taking a $q$th-power necessarily reduces the
singularity of $X_{u,v}.$ Also, we assume that the probability density
function of $X_{u,v}$ is skewed in the positive direction. This yields that $%
K\left( q\right) >0$ for $q>1.$ These assumptions, in conjunction with the
strict convexity of $K\left( q\right) ,$ suggest the assumption that 
\begin{equation}
K\left( 1\right) =0.  \label{2.6}
\end{equation}
This implies that 
\begin{equation}
EX_{u,v}=1\text{ for arbitrary }0<u<v.  \label{2.7}
\end{equation}

In this paper, we will consider smoothing at discrete scales $%
r_{j}=2^{-j+1}, $ $j=0,1,2,3,...$ Then the smoothed process at scale $r_{j}$
is 
\begin{equation}
X_{j}\left( t\right) =\varepsilon _{r_{j}}\left( t\right) =\frac{1}{2^{-j+1}}%
\int_{t-2^{-j}}^{t+2^{-j}}\varepsilon \left( s\right) ds.  \label{2.8}
\end{equation}
Under the condition $E\varepsilon \left( t\right) =1,$ it is reasonable to
assume that 
\begin{equation}
X_{0}\left( t\right) =1,\;t\in \left[ 0,1\right] .  \label{2.9}
\end{equation}
Then, at generation $J,$%
\begin{eqnarray}
X_{J}\left( t\right) &=&X_{0}\left( t\right) \frac{X_{1}\left( t\right) }{%
X_{0}\left( t\right) }\frac{X_{2}\left( t\right) }{X_{1}\left( t\right) }%
\,...\,\frac{X_{J}\left( t\right) }{X_{J-1}\left( t\right) }  \notag \\
&=&\frac{X_{1}\left( t\right) }{X_{0}\left( t\right) }\frac{X_{2}\left(
t\right) }{X_{1}\left( t\right) }\,...\,\frac{X_{J}\left( t\right) }{%
X_{J-1}\left( t\right) }.  \label{2.10}
\end{eqnarray}
Under the scale invariance conditions (i) and (ii), the random variables $%
X_{j}/X_{j-1}$ of (\ref{2.10}) are independent and have the same probability
distribution. Let $W$ denote a generic member of this family. Note that $%
EW=1 $ from (\ref{2.7}). Then (\ref{2.10}) can be rewritten as 
\begin{eqnarray}
X_{J}\left( t\right) &=&X_{J-1}\left( t\right) \frac{X_{J}\left( t\right) }{%
X_{J-1}\left( t\right) }  \notag \\
&=&W_{1}\left( t\right) W_{2}\left( t\right) ...W_{J}\left( t\right) ,\;t\in 
\left[ 0,1\right] .  \label{2.11}
\end{eqnarray}
In other words, $X_{J}\left( t\right) $ is a multiplicative cascade process
(see Holley and Waymire \cite{HW92}, Gupta and Waymire \cite{GW93}). Denote
by $\mu _{J}$ the sequence of random measures defined by the density $%
X_{J}\left( t\right) ,$ that is, 
\begin{equation*}
\mu _{J}\left( dt\right) =X_{J}\left( t\right) dt,\;J=1,2,3,...
\end{equation*}
It can be checked that $\mu _{J}$ \emph{a.s.} has a weak* limit $\mu
_{\infty }$ since for each bounded continuous function $f$ on $\left[ 0,1%
\right] ,$ the sequence $\int_{\left[ 0,1\right] }fd\mu _{J}$ is an $L_{1}$%
-bounded martingale (see Holley and Waymire \cite{HW92}, Mandelbrot \cite
{Man74}, Kahane and Peyriere \cite{KP76}). We denote the density
corresponding to $\mu _{\infty }$ by $X_{\infty }\left( t\right) .$ Then it
is seen from (\ref{2.8}) that 
\begin{equation}
X_{\infty }\left( t\right) =\varepsilon \left( t\right) ,\;t\in \left[ 0,1%
\right] .  \label{2.12}
\end{equation}
Summarising, we have established that 
\begin{eqnarray*}
&&\emph{\ The\ positive\ stationary\ process\,\,}\varepsilon \left( t\right)
\,\emph{is\ the\ limit\ of\ a\ } \\
&&\emph{\ multiplicative\ cascade\ with\ generator\ }W\emph{.}
\end{eqnarray*}

We next want to characterise this random cascade. We first note that, for $%
j=1,2,3,...,$%
\begin{equation}
\frac{X_{j}}{X_{j-1}}=2\frac{\int_{t-2^{-j}}^{t+2^{-j}}\varepsilon \left(
s\right) ds}{\int_{t-2^{-\left( j-1\right) }}^{t+2^{-\left( j-1\right)
}}\varepsilon \left( s\right) ds}\leq 2  \label{2.13}
\end{equation}
from the positivity of $\varepsilon \left( t\right) .$ Thus, 
\begin{equation*}
E\left( \frac{X_{j}}{X_{j-1}}\right) ^{q}\leq 2^{q}.
\end{equation*}
This inequality together with (\ref{2.4}) imply 
\begin{equation}
K\left( q\right) \leq q,\quad q\geq 0.  \label{2.14}
\end{equation}
We then have 
\begin{equation*}
\sum_{q=0}^{\infty }\left( E\left( \frac{X_{j}}{X_{j-1}}\right) ^{2q}\right)
^{-\frac{1}{2q}}=\sum_{q=0}^{\infty }\left( \frac{1}{2}\right) ^{\frac{%
K\left( 2q\right) }{2q}}=\infty .
\end{equation*}
In other words, the Carleman condition is satisfied (see Feller \cite{Fel71}%
, p. 224). As a result, we get 
\begin{eqnarray*}
&&T\emph{he\ probability\ density\ function\ }f_{W}\emph{\ of\ the\
generator\ }W\emph{\ is\ \ } \\
&&\emph{uniquely\,\,determined\ by\ the\ set\ }\left\{ K\left( q\right)
,\;q=0,1,2,...\right\} .
\end{eqnarray*}

It is seen that, if the function $K\left( q\right) $ has analytic
continuation into the complex plane, then the characteristic function of $%
\ln W$ has the form 
\begin{equation}
\psi \left( x\right) =E\left( e^{ix\ln W}\right) =\left( \frac{1}{2}\right)
^{-K\left( ix\right) }.  \label{2.15}
\end{equation}
Define $\psi _{n}\left( x\right) =\left( 1/2^{1/n}\right) ^{-K\left(
ix\right) }$ for an arbitrary integer $n.$ Then $\psi _{n}$ is the
characteristic function of the probability distribution corresponding to
smoothing with scales $\left( 2^{1/n}\right) ^{-j+1}.$ Also, it holds that 
\begin{equation*}
\psi \left( x\right) =\left( \psi _{n}\left( x\right) \right) ^{n}.
\end{equation*}
Thus $\psi \left( x\right) $ is infinitely divisible (see Feller \cite{Fel71}%
, p. 532); in other words, 
\begin{equation}
\ln W\,\,\text{\emph{has an infinitely divisible distribution.}}
\label{2.16}
\end{equation}
It is noted from (\ref{2.13}) that $-\ln \frac{W}{2}\geq 0.$ The most
general form for the characteristic function $\varphi \left( x\right) $ of
positive random variables is given by 
\begin{equation}
\varphi \left( x\right) =\exp \left\{ \int_{0}^{\infty }\frac{1-e^{ixs}}{s}%
P\left( ds\right) +iax\right\} ,  \label{2.17}
\end{equation}
where $a\geq 0$ and $P$ is a measure on the open interval $\left( 0,\infty
\right) $ such that $\int_{0}^{\infty }\left( 1+s\right) ^{-1}P\left(
ds\right) <\infty $ (see Feller \cite{Fel71}, p. 539). On the other hand, it
follows from (\ref{2.2}) and (\ref{2.4}) that the characteristic function of 
$-\ln \frac{W}{2}$ is given by 
\begin{eqnarray}
E\left( e^{-ix\ln \frac{W}{2}}\right) &=&2^{ix}E\left( W\right) ^{-ix} 
\notag \\
&=&2^{ix}\left( \frac{1}{2}\right) ^{-K\left( -ix\right) }.  \label{2.18}
\end{eqnarray}
Using $q=-ix$ and equating (\ref{2.17}) with (\ref{2.18}) then yields 
\begin{equation}
K\left( q\right) =\left( 1-\frac{a}{\ln 2}\right) q-\int_{0}^{\infty }\frac{%
1-e^{-qs}}{s}\frac{P\left( ds\right) }{\ln 2}.  \label{2.19}
\end{equation}
As constrained by (\ref{2.6}), the following condition must be satisfied by
the measure $P\left( ds\right) :$%
\begin{equation}
\int_{0}^{\infty }\frac{1-e^{-s}}{s}\frac{P\left( ds\right) }{\ln 2}=1-\frac{%
a}{\ln 2}\leq 1.  \label{2.20}
\end{equation}
Equations (\ref{2.19}) and (\ref{2.20}) provide the most general form for
the $K\left( q\right) $ curve of the positive random process $\left\{
\varepsilon \left( t\right) ,\;0\leq t\leq 1\right\} .$

In practice, fitting this $K\left( q\right) $ curve to data requires a
proper choice of the measure $P\left( ds\right) .$ Novikov \cite{Nov94}
suggests the use of the Gamma density function, namely, 
\begin{equation}
f\left( x\right) =Ax^{\alpha -1}\exp \left( -x/\sigma \right) ,  \label{2.21}
\end{equation}
where $P\left( dx\right) =f\left( x\right) dx$ and $A,\alpha ,\sigma $ are
positive constants. From (\ref{2.19}) and (\ref{2.21}) we get 
\begin{equation}
K\left( q\right) =\left\{ 
\begin{array}{ll}
\kappa \left( q-\frac{\left( q\sigma +1\right) ^{1-\alpha }-1}{\left( \sigma
+1\right) ^{1-\alpha }-1}\right) , & \alpha \neq 1, \\ 
\kappa \left( q-\frac{\ln \left( q\sigma +1\right) }{\ln \left( \sigma
+1\right) }\right) , & \alpha =1.
\end{array}
\right.  \label{2.22}
\end{equation}
where $\kappa =1-a/\ln 2$, and from (\ref{2.20}) we have
$$ A=\frac{\kappa \ln 2}{\sigma^{\alpha -1}\Gamma(\alpha -1)}(1-(\sigma+1)^{1-\alpha})^{-1}.
$$

The form (\ref{2.22}) will be used for data fitting in this paper. It is
seen from (\ref{2.2}) and (\ref{2.4}) that the data for the $K\left(
q\right) $ curve is provided by 
\begin{equation}
K\left( q\right) =\underset{J\rightarrow \infty }{\lim }\,\,\frac{\ln
E\left( X_{J}^{q}\right) }{-\ln 2^{-J+1}},  \label{2.23}
\end{equation}
where it should be noted from (\ref{2.12}) that $X_{\infty }\left( t\right)
=\varepsilon \left( t\right) ,$ the given positive random process.

Since each smoothed process $X_{J}$ may possess long-range dependence (see
Anh \emph{et al. }\cite{AHT99}), the ergodic theorem may not hold for these
processes. As a result, the computation of $E\left( X_{J}^{q}\right) $ as
sample averages may not be sufficiently accurate. There is an alternative
form of the ergodic theorem developed by Holley and Waymire \cite{HW92} for
random cascades which we now summarise.

For random cascades with density $\varepsilon \left( t\right) ,$ limit
measure $\mu _{\infty },$ branching number $b$ and generator $W,$ define 
\begin{equation}
M_{J}\left( q\right) =\sum_{k}^{{}}{}^{\prime }\left( \mu _{\infty }\left(
\Delta _{k}^{J}\right) \right) ^{q},  \label{2.24}
\end{equation}
\begin{equation}
\tau \left( q\right) =\underset{J\rightarrow \infty }{\lim }\,\frac{\ln
M_{J}\left( q\right) }{J\ln b},  \label{2.25}
\end{equation}
\begin{equation}
 D_q=\tau(q)/(q-1), \label{2.25a}
\end{equation}
\begin{equation}
\chi _{b}\left( q\right) =\log _{b}E\left( W^{q}\right) -\left( q-1\right) ,
\label{2.26}
\end{equation}
where the prime in (\ref{2.24}) indicates a sum over those subintervals $%
\Delta _{k}^{J}$ of generation $J$ which meet the support of $\mu _{\infty
} $.

\begin{theorem}
(Holley and Waymire \cite{HW92}) Assume that $W>a$ for some $a>0$ and $W<b$
with probability 1, and that $E\left( W^{2q}\right) /\left( EW^{q}\right)
^{2}<b.$ Then, with probability 1, 
\begin{equation}
\tau \left( q\right) =-\chi _{b}\left( q\right) .  \label{2.27}
\end{equation}
\end{theorem}

In our case as developed above, $b=2,$ and (\ref{2.13}) gives $W\leq 2$. In
fact the scale $r_{j}=2^{-j+1}$ used in (\ref{2.8}) is arbitrary; it can be $%
b^{-j+1}$ and the inequality $W\leq b$ still holds by definition of the
smoothing and the positivity of $\varepsilon \left( t\right) .$ In our
development, 
\begin{eqnarray*}
-K\left( q\right) &=&\underset{J\rightarrow \infty }{\lim }\,\,\frac{\ln
E\left( X_{J}^{q}\right) }{\ln 2^{-J+1}} \\
&=&\underset{J\rightarrow \infty }{\lim }\,\,\frac{J\ln E\left( W^{q}\right) 
}{\left( J-1\right) \ln 2^{-1}}\,\,\,\,\,\text{using (\ref{2.11})} \\
&=&-\frac{\ln E\left( W^{q}\right) }{\ln 2}.
\end{eqnarray*}
Consequently, 
\begin{equation}
K\left( q\right) =-\tau \left( q\right) +q-1.  \label{2.28}
\end{equation}
The above formula then provides a way to compute $K\left( q\right) $ via (%
\ref{2.25}) and (\ref{2.28}) using sums of $q$-th powers of the limit
measure instead of (\ref{2.23}) using expectations. In fact, the ergodic
theorem now takes the following form 
\begin{equation*}
\underset{J\rightarrow \infty }{\lim }\frac{\ln E\left( X_{J}^{q}\right) }{%
\left( J-1\right) \ln 2}=\underset{J\rightarrow \infty }{\lim }\frac{\ln
\sum_{k}^{\prime }\left( \mu _{\infty }\left( \Delta _{k}^{J}\right) \right)
^{q}}{J\ln 2}+q-1.
\end{equation*}

\section{Data fitting and discussion}

 \ \ \ \ For $K=8$, we first calculated $K(q)$
of the measure representation of all the above organisms directly from the
definition of $K(q)$ (\ref{2.23}). Figure 2 shows how to calculate this $K(q)$ curve. We
give the $K(q)$ curves of \textit{E. coli, S. cerevisiae Chr15, C. elegans
Chr1, A. thaliana Chr2,} and \textit{Homo sapiens Chr22} in Figure 3. From
Figure 3, it is seen that the grade of the organism is lower when the $K_q$ curve is
flatter. Hence the evolution relationship of these organisms is apparent. We
denote by $K_{d}(q)$ the value of $K(q)$ computed from the data using its
definition (\ref{2.23}) and define 
\begin{equation*}
error=\sum_{j=1}^{J}|\kappa (q_{j}-\frac{(q_{j}\sigma +1)^{1-\alpha }-1}{%
(\sigma +1)^{1-\alpha }-1})-K_{d}(q_{j})|^{2}.
\end{equation*}
Then the values of $\kappa ,\,\sigma $ and $\alpha $ can be estimated
through minimising $error$. In this minimisation, we assume 
\begin{equation*}
0\leq \kappa ,\ \sigma ,\ \alpha \leq 20.
\end{equation*}
After obtaining the value of $\kappa $, $\sigma $ and $\alpha $, we then get
the $K(q)$ curve from (\ref{2.22}). The data fitting based on the form (\ref
{2.22}) was performed on all the organisms and shown in Table 1 (from top to
bottom, in the increasing order of the value of $\kappa ).$ It is found that
the form (\ref{2.22}) gives a perfect fit to the data for all bacteria. As an example, we
give the data fitting of \textit{E. coli, S. cerevisiae Chr15} and \textit{%
C. elegans Chr1} in Figure 4.   But for higher organisms, for example,
{\it Homo sapiens Chromosome 22}, the fitting is not as good.
Note that we only selected one chromosome for each higher organism. If 
all chromosomes for each higher organism are considered, 
the data fitting for $K_q$ will be
better. The fit for Human Chromosome is the worst in Table 1. Since
the length of Human Chromosome 22 is not larger than those of the  complete
genomes of all bacteria, there does not seem to be any relationship between 
the quality of fit and the length of the complete genome.

   The parameter $\kappa$ provides a tool to classify bacteria.  From Table 1,
 one can see {\it Helicobacter pylori 26695} and {\it Helicobacter pylori J99}
 group together, and three Chlamydia almost group together. 
 But this parameter $\kappa$ alone is not sufficient, it must be combined
 with other tools to classify bacteria.

We also calculated the values of $\tau (q)$ using its definition (\ref{2.25}%
). We found the values of $K_{d}(q)$ coincide with those obtained from (\ref
{2.28}). Hence we indeed can use (\ref{2.28}) to calculate $K(q)$.
Formula (\ref{2.22}) gives an analytical expression for the quantity $K_q$. 
An analytical expressions for $\tau(q)$ can therefore be obtained 
from (\ref{2.28})
and $D_q$ from (\ref{2.25a}).

\section{ Conclusions}

\ \ \ \ The idea of our measure representation is similar to the portrait method
proposed by Hao \textit{et al.}\cite{HLZ00}. It provides a simple yet
powerful visualisation method to amplify the difference between a DNA
sequence and a random sequence as well as to distinguish DNA sequences
themselves in more details. From our measure representation we can exactly 
know the frequencies of all the $K$-string appearing in the complete genome.
But the representations alone are not sufficient to discuss the classification
and evolution problem. Hence we need further tools.

In our previous work (Yu {\it et al} \cite{YA00b}), when the measure
representations of organisms were viewed as time series, it was found that
they are far from being random time series, and in fact exhibit strong
long-range dependence. Multifractal analysis of the complete genomes was
performed in relation to the problem of classification and evolution of
organisms. In this paper, we established a theoretical model of the
probability distribution of the complete genomes. This probability
distribution, particularly the resulting $K(q)$ curve, provides a precise
tool for their characterisation. Numerical results confirm the accuracy of
the method of this paper.

For a completely random sequence based on the alphabet $\{a,c,g,t\}$, we have
$D_q=1,\ \tau(q)=q-1,\ K(q)=0$ for all $q$. From the $K(q)$ curves,
it is seen that all complete genomes selected are far from being a completely
 random sequence.

\section*{Acknowledgement}
  
  \ \ \ \ The Authors would like to express their gratitude to the referees for 
good comments and suggestions to improve this paper. This research was partially
supported by QUT Postdoctoral Research Grant 9900658 to Zu-Guo Yu, and the
RGC Earmarked Grant CUHK 4215/99P.

\begin{table}[tbp]
\caption{The values of $\kappa ,\,\sigma ,\,\alpha ,\,error$ of all
the organisms selected.}
\label{tableksa}
\begin{center}
\begin{tabular}{|l|c|c|c|c|}
\hline
\ \ \ \ \ \ Species & $\kappa$ & $\sigma$ & $\alpha$ & $error$ \\ \hline
Aquifex aeolicus & 0.210967 & 0.034741 & 20.000000 & 1.149058E-03 \\ 
Haemophilus influenzae & 0.250405 & 0.026628 & 20.000000 & 1.718141E-04 \\ 
Synechocystis sp. PCC6803 & 0.252695 & 0.023009 & 14.895300 & 2.551734E-04
\\ 
Mycoplasma pneumoniae & 0.260598 & 0.028227 & 14.468067 & 1.367545E-04 \\ 
Chlamydia pneumoniae AR39 & 0.261441 & 0.015080 & 20.000000 & 1.109025E-02
\\ 
Rhizobium sp. NGR234 & 0.269307 & 0.141332 & 1.974406 & 1.037725E-05 \\ 
Chlamydia muridarum & 0.282757 & 0.021999 & 20.000000 & 5.528718E-03 \\ 
Chlamydia trachomatis & 0.285242 & 0.016422 & 20.000000 & 3.569117E-03 \\ 
Neisseria meningitidis MC58 & 0.287688 & 0.021525 & 20.000000 & 3.869811E-04
\\ 
Helicobacter pylori 26695 & 0.296316 & 0.042743 & 20.000000 & 3.999003E-03
\\ 
Helicobacter pylori J99 & 0.300842 & 0.039837 & 20.000000 & 4.532450E-03 \\ 
Methanococcus jannaschii & 0.305624 & 0.034413 & 19.737016 & 9.356220E-05 \\ 
Rickettsia prowazekii & 0.312790 & 0.036216 & 19.484758 & 1.681558E-04 \\ 
Neisseria meningitidis Z2491 & 0.316484 & 0.021405 & 20.000000 & 4.444530E-04
\\ 
Bacillus subtilis & 0.325036 & 0.015238 & 20.000000 & 5.327829E-03 \\ 
Aeropyrum pernix & 0.325043 & 0.024461 & 20.000000 & 1.056628E-02 \\ 
Mycoplasma genitalium & 0.326433 & 0.033756 & 20.000000 & 1.517762E-03 \\ 
Campylobacter jejuni & 0.342793 & 0.044513 & 20.000000 & 1.316877E-03 \\ 
M. tuberculosis & 0.345510 & 0.020729 & 19.509203 & 4.187475E-04 \\ 
Borrelia burgdorferi & 0.350140 & 0.045101 & 20.000000 & 2.282837E-03 \\ 
Thermotoga maritima & 0.364864 & 0.017640 & 20.000000 & 1.094542E-03 \\ 
Treponema pallidum & 0.365539 & 0.011555 & 20.000000 & 7.890963E-03 \\ 
Ureaplasma urealyticum & 0.371367 & 0.067125 & 12.859609 & 2.250143E-04 \\ 
Escherichia coli & 0.386280 & 0.024556 & 6.404487 & 2.418786E-04 \\ 
M. thermoautotrophicum & 0.388544 & 0.015769 & 13.884240 & 1.474283E-03 \\ 
Pseudomonas aeruginosa & 0.412200 & 0.753456 & 0.918436 & 4.798280E-05 \\ 
Caenorhabditis elegans Chr1 & 0.440354 & 0.030755 & 20.000000 & 1.087368E-02
\\ 
Chlamydia pneumoniae AR39 & 0.484163 & 0.018637 & 20.000000 & 2.701796E-03
\\ 
Archaeoglobus fulgidus & 0.487055 & 0.016984 & 11.046987 & 1.435727E-03 \\ 
S. cerevisiae Chr15 & 0.511099 & 0.014271 & 11.487615 & 2.237813E-03 \\ 
Pyrococcus abyssi & 0.513144 & 0.016623 & 7.295978 & 7.294311E-04 \\ 
Buchnera sp. APS & 0.536577 & 0.031866 & 20.000000 & 4.064171E-03 \\ 
Arabidopsis thaliana Chr2 & 0.546252 & 0.014951 & 13.096780 & 2.629544E-03
\\ 
Pyrococcus abyssi & 0.562316 & 0.015389 & 11.328229 & 1.574777E-03 \\ 
Vibrio cholerae & 0.604051 & 0.028218 & 3.209793 & 3.147899E-04 \\ 
Plasmodium falciparum Chr3 & 0.769704 & 0.049365 & 20.000000 & 4.257000E-02 \\ 
Xylella fastidiosa & 1.014092 & 0.010085 & 7.503579 & 1.194219E-02 \\ 
Homo sapiens Chr22 & 1.290643 & 0.008267 & 12.96619 & 1.900450E-01 \\ 
\hline
\end{tabular}
\end{center}
\end{table}

\begin{figure}[tbp]
\centerline{\epsfxsize=8cm \epsfbox{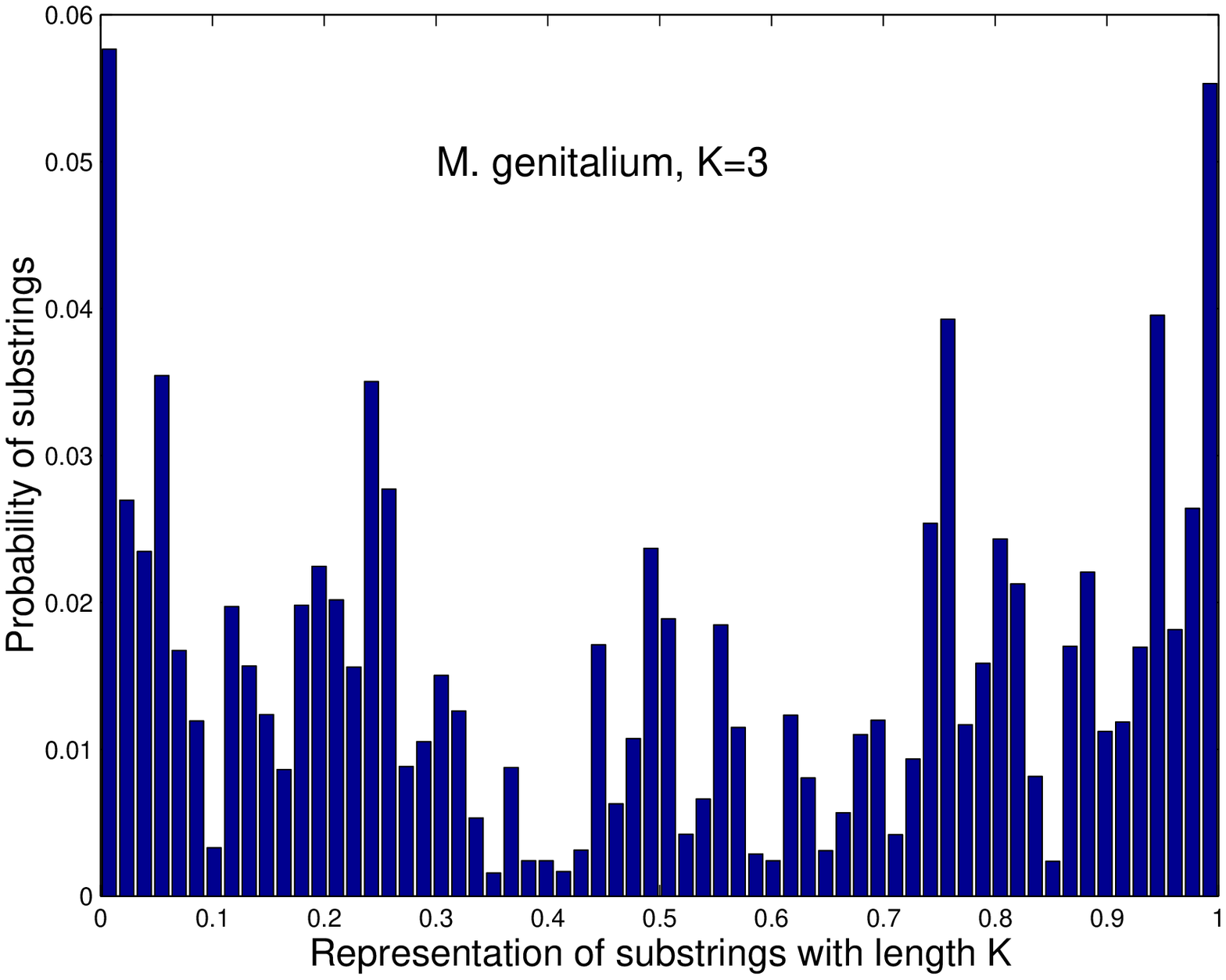}
\epsfxsize=8cm \epsfbox{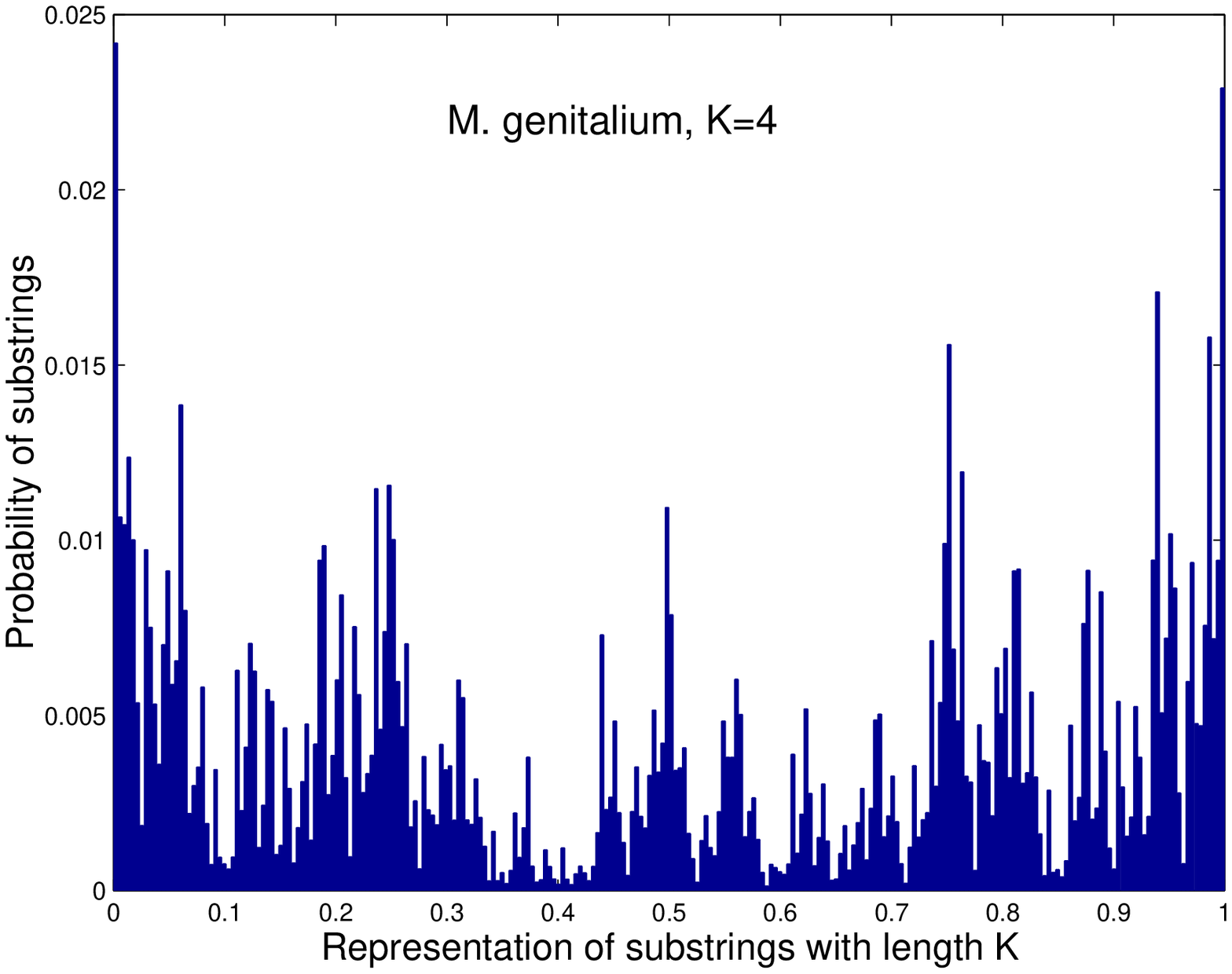}} 
\centerline{\epsfxsize=8cm \epsfbox{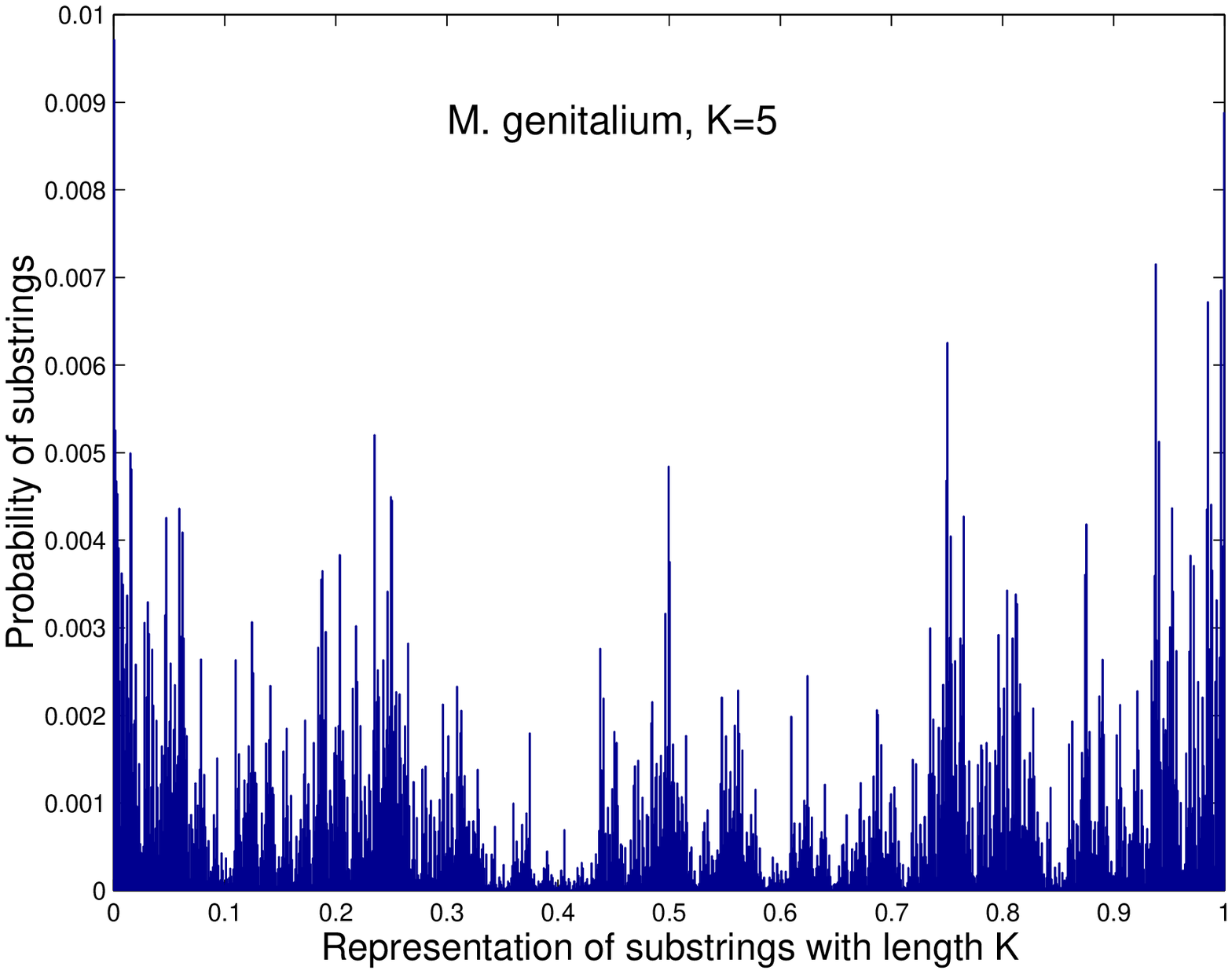}
\epsfxsize=8cm \epsfbox{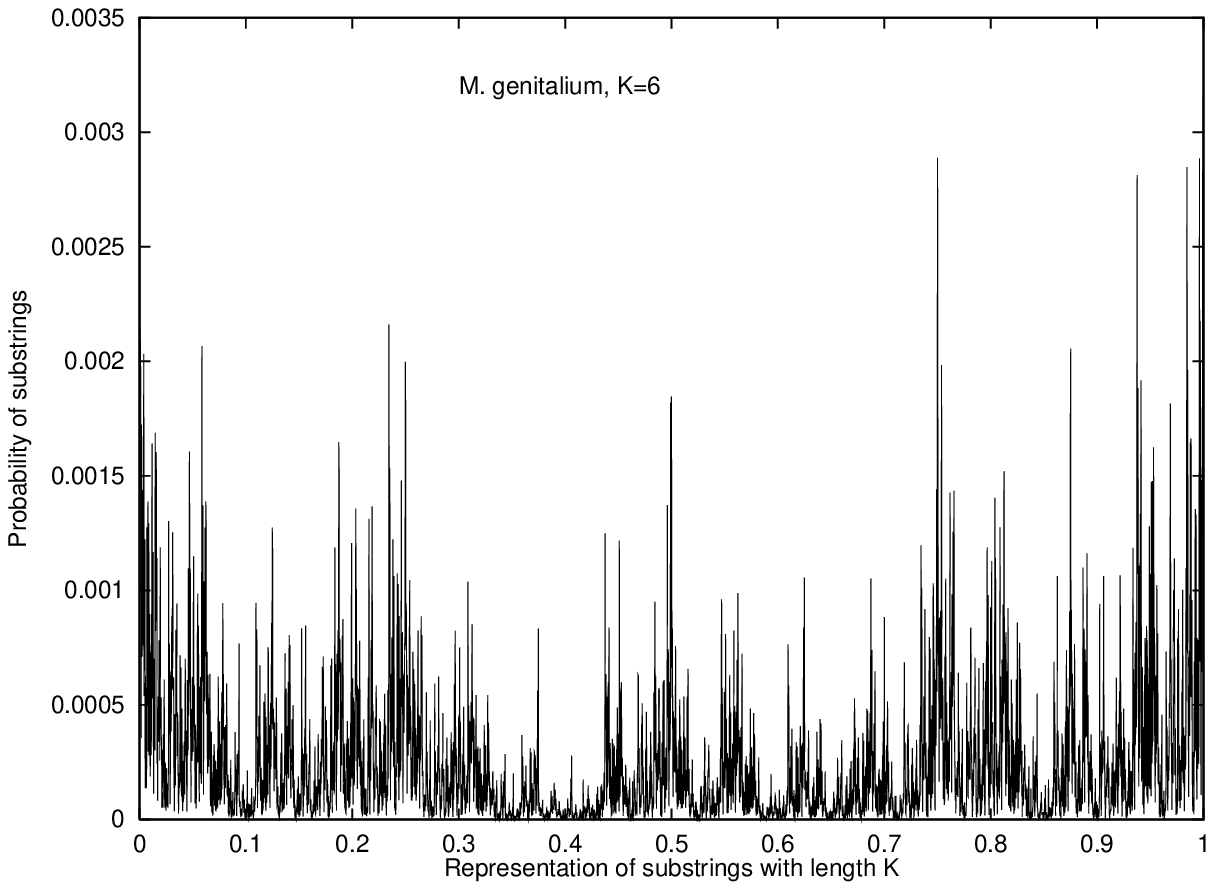}} 
\centerline{\epsfxsize=8cm \epsfbox{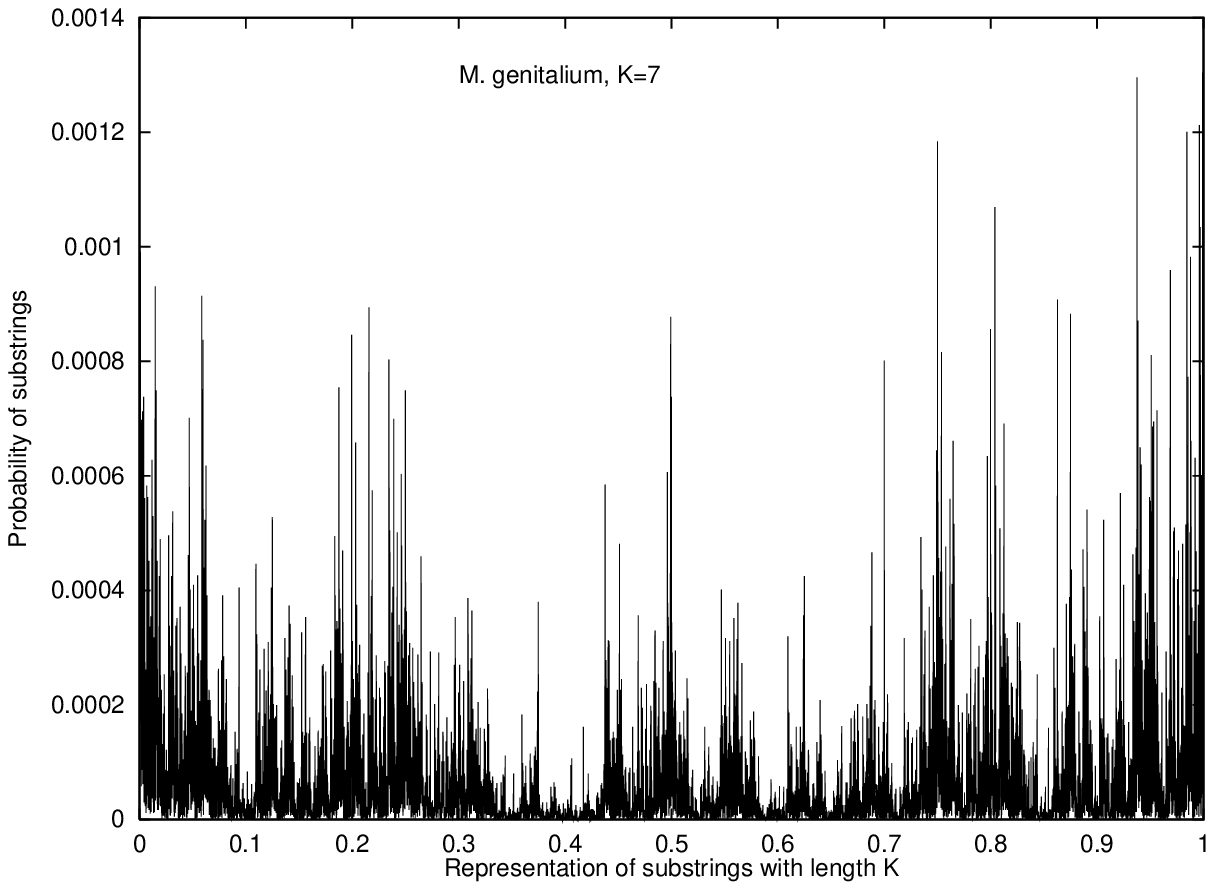}
\epsfxsize=8cm \epsfbox{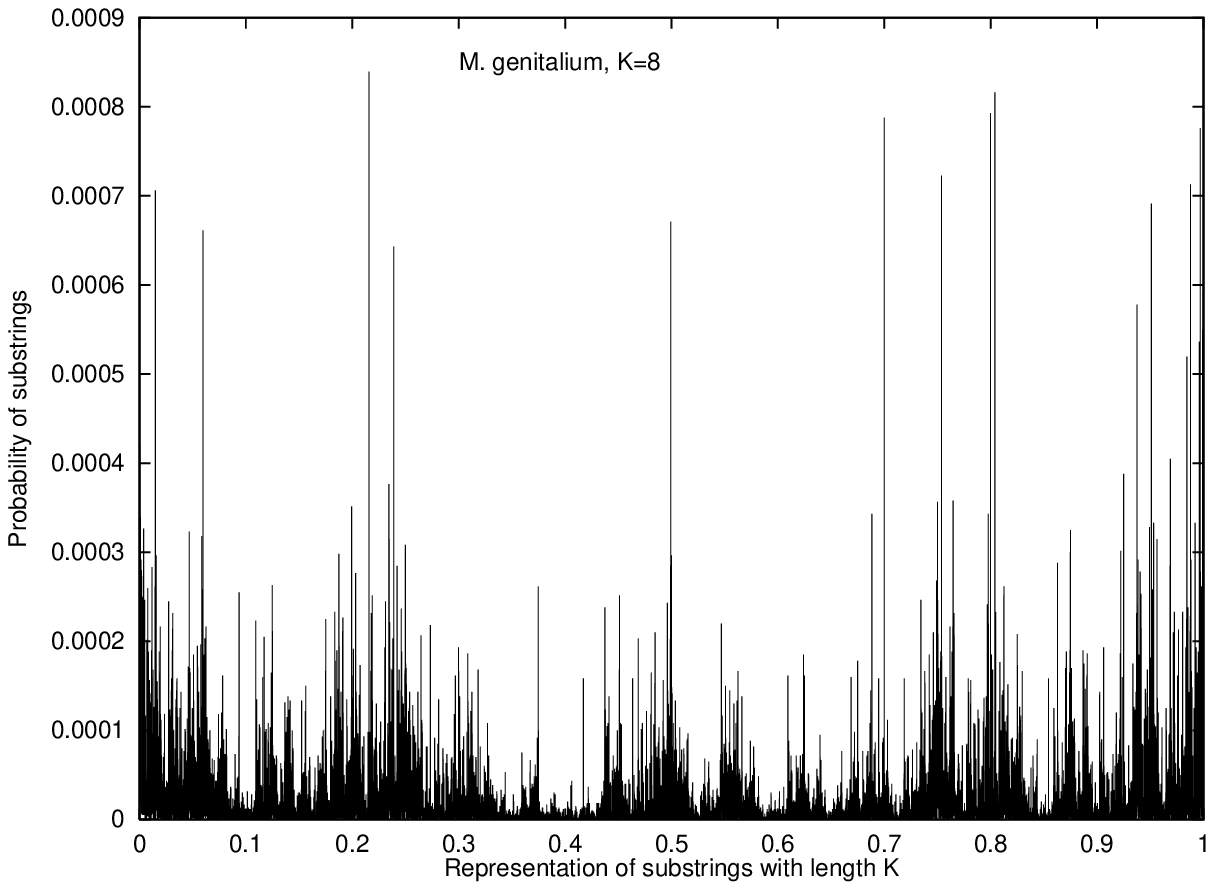}}
\caption{{\protect\footnotesize Histograms of substrings with different
lengths}}
\label{mgencd}
\end{figure}

\begin{figure}[tbp]
\centerline{\epsfxsize=12.5cm \epsfbox{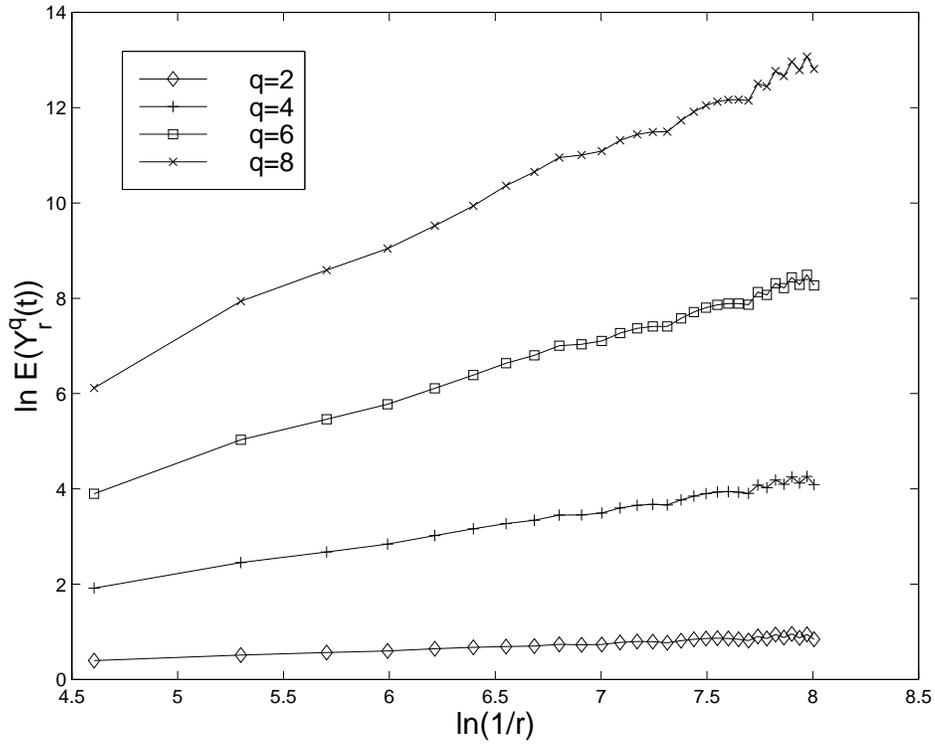}
}
\caption{{\protect\footnotesize An example to show how to obtain the value
of $K(q)$ directly using its definition.}}
\label{linecal}
\end{figure}

\begin{figure}[tbp]
\centerline{\epsfxsize=11.5cm \epsfbox{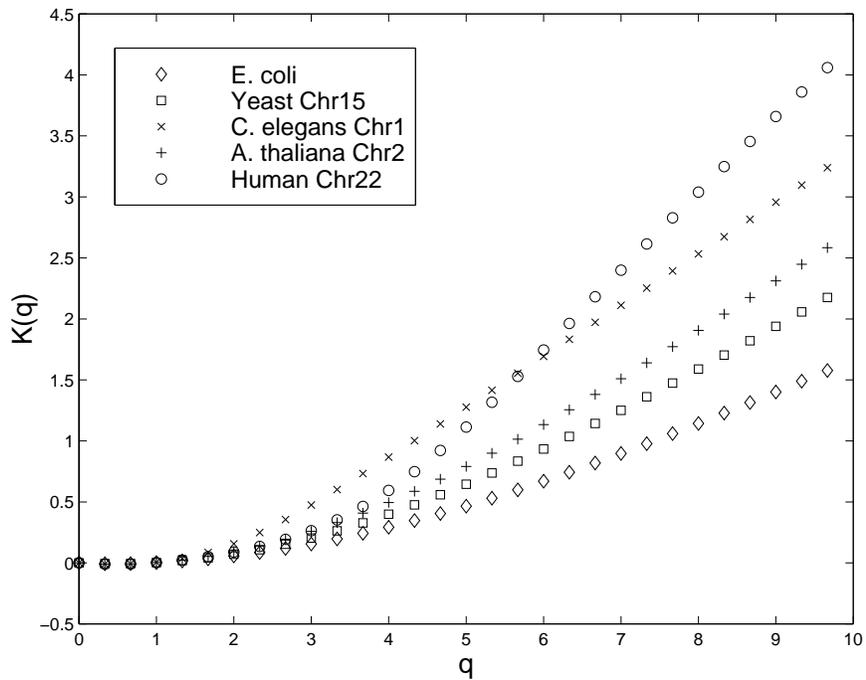}
}
\caption{{\protect\footnotesize The values of $K(q)$ of Chromosome 22 of
Homo sapiens, Chromosome 2 of A. thaliana, Chromosome 1 of C. elegans,
Chromosome 15 of S. cerevisiae and E. coli. }}
\label{kqevolu}
\end{figure}

\begin{figure}[tbp]
\centerline{\epsfxsize=12.5cm \epsfbox{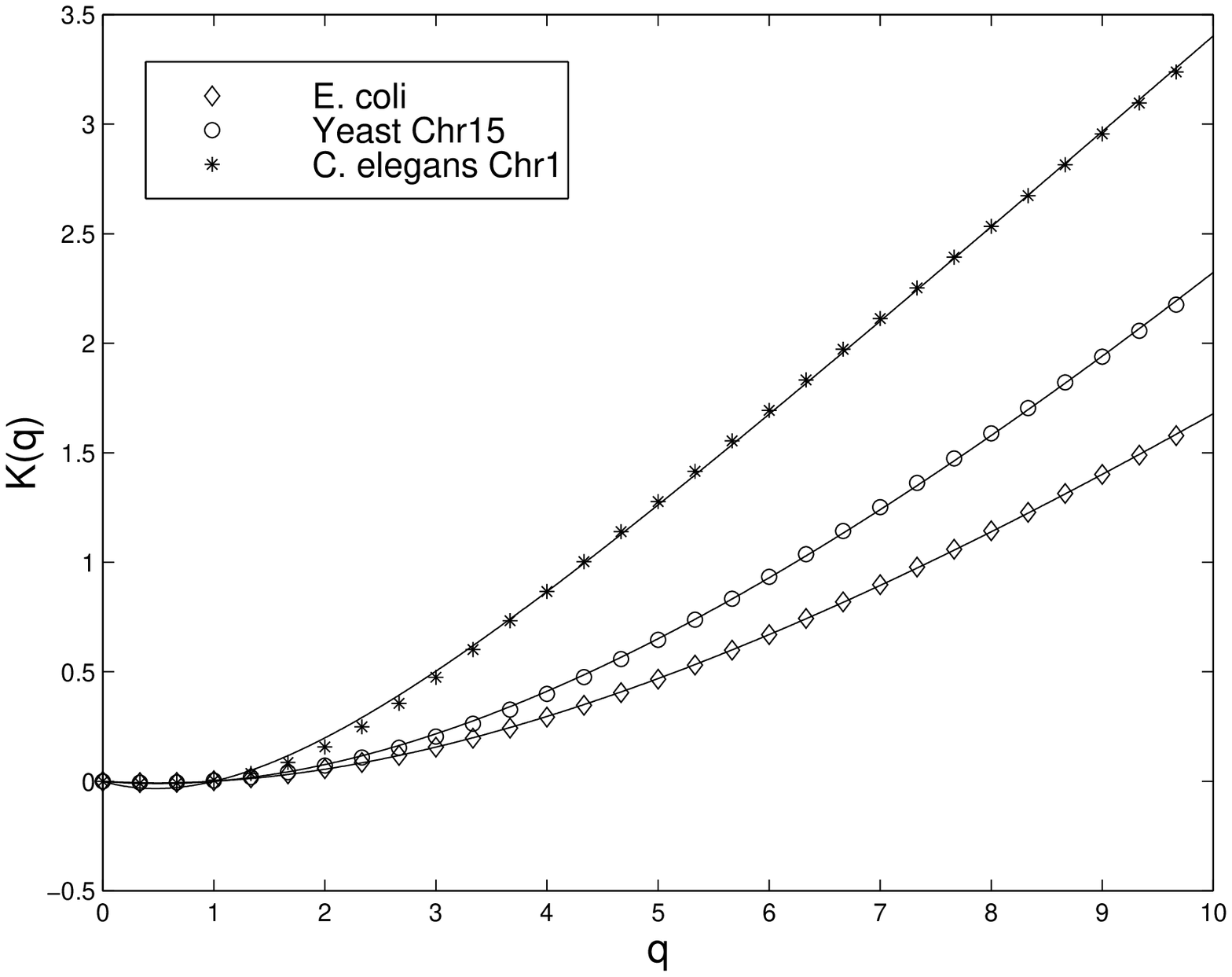}
}
\caption{{\protect\footnotesize The data fitting of E. coli,chromosome 15 of S. cerevisiae and
chromosome 1 of C. elegans based on the Gamma model. The symbolled curves
represent $K_{d}\left( q\right) $ computed from data, while the continuous
curves represent $K\left( q\right) $ computed from formula (\ref{2.22}).}}
\label{kqfitting}
\end{figure}


\bigskip

\begin{thebibliography}{10}
\renewcommand{\baselinestretch}{0.05} \parskip=-0.2mm
{\small 

\bibitem{AHT99}
V.~V. Anh, C.~C. Heyde, and Q.~Tieng.
\newblock Stochastic models for fractal processes.
\newblock {\em Journal of Statistical Planning and Inference},
  80(1/2):123--135, 1999.

\bibitem{BGR94}
C.~L. Berthelsen, J.~A. Glazier, and S.~Raghavachari.
\newblock {\em Phys. Rev. E}, 49(3):1860, 1994.

\bibitem{Vie99}
M.~de~Sousa~Vieira.
\newblock Statistics of {DNA} sequences: A low-frequency analysis.
\newblock {\em Phys. Rev. E}, 60(5):5932--5937, 1999.

\bibitem{Fel71}
W.~Feller.
\newblock {\em An Introduction to Probability Theory and its Applications},
  volume~II.
\newblock Wiley, New York, 1971.

\bibitem{Gol93}
N.~Goldman.
\newblock Nucleotide, dinucleotide and trinucleotide frequencies explain
  patterns observed in chaos game representations of {DNA} sequences.
\newblock {\em Nucleic Acids Research}, 21(10):2487--2491, 1993.

\bibitem{GW93}
V.~K. Gupta and E.~C. Waymire.
\newblock A statistical analysis of mesoscale rainfall as a random cascade.
\newblock {\em Journal of Applied Meteorology}, 32:251--267, 1993.

\bibitem{HLZ00}
B.-L. Hao, H.-C. Lee, and S.-Y. Zhang.
\newblock Fractals related to long {DNA} sequences and complete genomes.
\newblock {\em Chaos, Solitons and Fractals}, 11(6):825--836, 2000.

\bibitem{HXYC01}
B.-L. Hao, H.-M. Xie, Z.-G. Yu, and G.-Y. Chen.
\newblock Avoided strings in bacterial complete genomes and a related
  combinatorial problem.
\newblock {\em Ann. of Combinatorics}, 4:247-255, 2000.


\bibitem{HW92}
R.~Holley and E.~C. Waymire.
\newblock Multifractal dimensions and scaling exponents for strongly bounded
  random cascades.
\newblock {\em The Annals of Applied Probability}, 2(4):819--845, 1992.

\bibitem{Jef90}
H.~J. Jeffrey.
\newblock Chaos game representation of gene structure.
\newblock {\em Nucleic Acids Research}, 18(8):2163--2170, 1990.

\bibitem{KP76}
J.-P. Kahane and J.~Peyri{\`e}re.
\newblock Sur certaines martingales de benoit mandelbrot.
\newblock {\em Advances in Mathematics}, 22:131--145, 1976.

\bibitem{Man74}
B.~B. Mandelbrot.
\newblock Intermittent turbulence in self-similar cascades: divergence of high
  moments and dimension of the carrier.
\newblock {\em J. Fluid Mech.}, 62:331--358, 1974.

\bibitem{Nov94}
E.~A. Novikov.
\newblock Infinitely divisible distributions in turbulence.
\newblock {\em Physical Review E}, 50(5), 1994.

\bibitem{PA00}
A.~Provata and Y.~Almirantis.
\newblock Fractal cantor patterns in the sequence structure of {DNA}.
\newblock {\em Fractals}, 8(1):15--27, 2000.

\bibitem{YA00b}
Z.-G. Yu, V.~Anh and K.-S. Lau.
\newblock Measure representation and multifractal analysis of complete genomes.
\newblock {\em Phys. Rev. E}, Sept. of 2001, (To appear).

\bibitem{YA00}
Z.-G. Yu and V.~Anh.
\newblock Time series model based on global structure of complete genome.
\newblock {\em Chaos, Solitons and Fractals},  12(10):1827-1834, 2001.

\bibitem{YAW00}
Z.-G. Yu, V.~V. Anh, and B.~Wang.
\newblock Correlation property of length sequences based on global structure of
  complete genome.
\newblock {\em Phys. Rev. E}, 63: 11903, 2001.

\bibitem{YHXC00}
Z.-G. Yu, B.-L. Hao, H.-M. Xie, and G.-Y. Chen.
\newblock Dimension of fractals related to language defined by tagged strings
  in complete genome.
\newblock {\em Chaos, Solitons and Fractals}, 11(14):2215--2222, 2000.

}
\end{thebibliography}

\end{document}